\documentclass[aps,twocolumn,pra,superscriptaddress,nofootinbib,amsmath,amssymb,floats,floatfix,english,a4paper]{revtex4-1}
\usepackage{polski}
\usepackage[utf8]{inputenc}
\usepackage[T1]{fontenc}
\usepackage{amsmath}
\usepackage{amssymb}
\usepackage{ifthen}
\usepackage{graphicx}
\usepackage{times}
\usepackage{babel}
\usepackage{color}
\usepackage{bm}
\usepackage{pstricks}

\usepackage[percent]{overpic}
\usepackage{tikz}
\usepackage{ulem}

\DeclareMathSizes{7}{7}{5}{4} 
\DeclareMathSizes{6}{6}{3}{2} 
\DeclareMathSizes{5}{5}{3}{2} 

\newcommand{\smaller}{\fontsize{7}{6}\selectfont}

\DeclareMathSizes{5}{5}{2}{2}

\newcommand{\op}[1]{\widehat{#1}}
\newcommand{\dagop}[1]{\widehat{#1}^{\dagger}}
\newcommand{\bo}[1]{{\mathbf{#1}}}
\newcommand{\mc}[1]{{\mathcal{#1}}}
\newcommand{\wt}[1]{{\widetilde{#1}}}
\newcommand{\wb}[1]{{\overline{#1}}}

\newcommand{\nonu}{\nonumber}
\newcommand{\etal}{~\textsl{et al.}}

\newlength{\templength}

\newenvironment{eq}{\begin{equation}}{\end{equation}}

\newcommand{\eqn}[1]{(\ref{#1})}

\renewcommand{\eq}[2]{\begin{equation}\label{#1}#2\end{equation}}
\newcommand{\eqs}[2]{\begin{subequations}\label{#1}\begin{eqnarray}#2\end{eqnarray}\end{subequations}}
\newcommand{\eqa}[2]{\begin{eqnarray}\label{#1}#2\end{eqnarray}}

\usepackage{lmodern}

\newcommand{\REM}[1]{\ifthenelse{0=1}{#1}{}}

\newcommand{\ve}{\varepsilon}

\begin{document}
\title{
Complex wave fields in the interacting 1d Bose gas
}

\author{J. Pietraszewicz}
\affiliation{Institute of Physics, Polish Academy of Sciences, Aleja Lotnik\'ow
32/46, 02-668 Warsaw, Poland}
\email{pietras@ifpan.edu.pl, deuar@ifpan.edu.pl}

\author{P. Deuar}
\affiliation{Institute of Physics, Polish Academy of Sciences, Aleja Lotnik\'ow
32/46, 02-668 Warsaw, Poland}

\date{\today}

\begin{abstract}
We study the temperature regimes of the 1d interacting gas to determine when the matter wave (c-field) theory is, in fact, correct and usable. The judgment is made by investigating the level of discrepancy in many observables at once in comparison to the exact Yang-Yang theory. We also determine what cutoff maximizes the accuracy of such an approach. Results are given in terms of a bound on accuracy, as well as an optimal cutoff prescription. For a wide range of temperatures the optimal cutoff is independent of density or interaction strength and so its temperature dependent form is suitable for many cloud shapes and, possibly, basis choices. However, this best global choice  is higher in energy than most prior determinations. The high value is needed to obtain the correct kinetic energy, but does not detrimentally affect other observables.
\end{abstract} 
\pacs{}

\maketitle 

%%%%%%%%%%%%%%%%%%%%%%%%%%%%%%%%%%%%%%%%%%%%%%%%%%%%%%%%%%%%%%%%%%%%%%%%%%%%%%%%%%%%%%%%%%%%%%%%%%%%%%%%%%%%%%%%%%%%%%%%%%%%%%%%%%%%%%%%%%%%%%%%%%%%%%%%%%%%%%%%%%%%%%%%%%%%%%%%%%%%%%%%%%%%%%%%%%%%%%%%%%%%%%%%%%%%%%%%%%%%%%%%%%%%%%%%%%%%
\section{Introduction}
\label{INTRO} 

 The classical field, or ``matter wave'', description has become an irreplaceable workhorse
 in the quantum gases community. It allows one to deal with a zoo of dynamical and thermal phenomena,
 particularly those that are non-perturbative or random between experimental runs. 
 The idea of the c-field approach is to treat the low energy part of the system as an ensemble of complex-valued wave fields 
 when mode occupations are large. The general aim here will be to quantitatively judge the accuracy of the c-field approach
 and circumscribe the physical parameters of the matter wave region.   
 This will make calculations more confident in the future.

 To say that a system's state is well described by such matter waves implies a host of important physical consequences. Among them: 
 the system behaves like a superfluid at least on short scales; 
 all relevant features come from a collective contribution of many particles;
 the capability for quantum wave turbulence and nonlinear self-organization\cite{Tsatsos16}; 
and single realizations of the ensemble correspond to measurements of many particle positions in 
 an experimental run \cite{Kagan97,Duine01,Lewis-Swan16}.

 A rapid growth of interest in these kinds of phenomena has occurred as soon as
 they became accessible experimentally. Their theoretical study and comparison to experiment 
 has relied on the various flavors of classical field descriptions \cite{FINESS-Book-Brewczyk,FINESS-Book-Davis,FINESS-Book-Cockburn}. 
 This is because c-fields are typically the only non-perturbative treatment of quantum mechanics that remains tractable for a large systems.
 Examples include defect seeding and formation 
\cite{Weiler08,Bradley08,Damski10,Witkowska11,Liu16}, 
quantum turbulence \cite{Berloff02,Parker05,Tsatsos16}, 
the Kibble-Zurek mechanism \cite{Sabbatini12,Swislocki13,Anquez16},
nonthermal fixed points \cite{Nowak12,Karl13} and vortex dynamics \cite{Bisset09a,Karpiuk09,Rooney10,Simula14}, 
the BKT transition \cite{Bisset09a}, and
evaporative cooling \cite{Marshall99,Proukakis06,Witkowska11}. 
Related effective field theories using complex-valued fields have been developed  for polaritons \cite{Wouters09,Chiocchetta14}, superfluid Fermi gases \cite{Lacroix13,Klimin15} or for Yang-Mills theory \cite{Moore97}.

 The general understanding of c-fields for the last 15 years or so has rested on two widely applicable qualitative arguments:
 Firstly, that the bulk of the physics must take place in single particle modes that are highly occupied.
 This allows one to neglect particle discretization and the non-commutativity of the Bose fields $\op{\Psi}(\bo{x})$ and $\dagop{\Psi}(\bo{x})$. 
 Secondly, that the subspace of modes treated with c-fields should be limited to below some energy cutoff $E_c$,
 at which mode occupation is $\mc{O}(1-10)$. At this level particle discretization
 (which is impossible to emulate using a complex amplitude)  becomes important.
 Such a cutoff also prevents the ultraviolet catastrophe that occurs
 because of the equipartition of $k_BT$ energy per each mode in classical wave equilibrium.
 A variety of prescriptions for choosing the cutoff have been obtained in the past 
\cite{Witkowska09,Brewczyk04,Zawitkowski04,Sinatra02,Rooney10,Bradley05,Blakie07,Sato12,Cockburn12},
 but they were usually tailored to correctly predict one particular observable. 

For a general treatment of a system with classical fields, however, one desires something broader:
 that at least all the usual measured observables are close to being described correctly at the same time.
 This is particularly important for the study of nonlinear dynamical and nonequilibrium phenomena
 such as quantum turbulence, or defect formation times, for which significant errors in one quantity
 will rapidly feed through into errors in all others. 
 As we will show, the key to a better understanding of the situation is that some quantities are much more sensitive
 to cutoff than others. Moreover, their dependence can come in two flavors
 depending on whether they are dominated by low energy (IR) or high energy (UV) modes.

 Past cutoff studies concentrated primarily on static observables dominated by low energy modes
 (condensate fraction \cite{Zawitkowski04,Blakie07,Witkowska09}, phase \cite{Sato12} and density correlation functions \cite{Cockburn12}).
 For damping and long-time dynamics, though \cite{Sinatra02,Sinatra08,Bezett09a,Karpiuk10}, the influence of higher energy modes comes out more strongly.
 A cutoff determination that would cover such nonequilibrium processes has been lacking.
 Here we will take the approach that to describe them well, at least the energy and its components (kinetic, interacting) should be correct. Accordingly, we will determine the cutoffs needed for both static and energetic observables to be accurate simultaneously, and
explicitly bound this accuracy.

 In this paper we consider the one-dimensional Bose gas with repulsive contact interactions.
 This system underpins a very large part of ultracold gas experiment and theory. 
 Moreover, there exist beautiful exact results for the uniform gas to compare to \cite{Lieb63,Yang69}. 
 To judge when matter wave physics is an accurate description, we generate c-field ensembles,
 construct a robust figure of merit
 and study its dependence on the cutoff. This follows the same tested route that was used
 in a preliminary study of the ideal Bose gas in \cite{Pietraszewicz15}. 
 
 An overview of the classical field method and system properties is given in Sec.~\ref{SETTING}.
 Next, Sec.~\ref{OBS} reports the dependence of observables on the cutoff, while 
 in Sec.\ref{PROCEDURE} the procedure used to judge the overall goodness of a c-field description is presented. 
 The practical limits of the matter wave region and the globally optimal cutoff are found in Sec.~\ref{RMS}.
 Subsequently, analysis of the optimal cutoff for nonuniform and/or dynamical systems can be found in Sec.\ref{ADDINFO},
 along with a comparison to earlier results. We conclude in Sec.~\ref{CONCLUSIONS}. 
Additional detail on a number of technical matters is given in the appendices.

%%%%%%%%%%%%%%%%%%%%%%%%%%%%%%%%%%%%%%%%%%%%%%%%%%%%%%%%%%%%%%%%%%%%%%%%%%%%%%%%%%%%%%%%%%%%%%%%%%%%%%%%%%%%%%%%%%%%%%%%%%%%%%%%%%%%%%%%%%%%%%%%%%%%%%%%%%%%%%%%%%%%%%%%%%%%%%%%%%%%%%%%%%%%%%%%%%%%%%%%%%%%%%%%%%%%%%%%%%%%%%%%%%%%%%%%%%%%%
\section{Description of the system}
\label{SETTING}

\subsection{Classical wave fields}
\label{CFIELDS}
The classical field description of a system can be succinctly summarized as the replacement of quantum annihilation (creation) operators  $\hat{a}_k$ ($\hat{a}_k^{\dagger}$)
 of single particle modes $k$ in the second-quantized field operator 
 by complex amplitudes  $\alpha_k$ ($\alpha^{*}_k$). With mode wave functions $\psi_k({\bf x})$, it can be written:
\eq{cfield}{
       \hat{\Psi}( {\bf x}) = \sum_{k} \hat{a}_k  \psi_k( {\bf x} ) 
       \to \Psi({\bf x}) = \Bigg{\{} \sum_{k\in \mathcal{C}}  \alpha_k  \psi_k( {\bf x}) \Bigg{\}}.
}
This is warranted when occupations are sufficiently macroscopic. 
Evidently, occupations will become \emph{not} sufficiently macroscopic for modes with high enough energy. For this reason it is necessary to restrict the set of modes to
a low energy subspace (often called the ``coherent'' or ``c-field'' region \cite{FINESS-Book-Davis}), which is denoted by $\mathcal{C}$. This set is usually parametrized by a single cutoff parameter, at a given mode energy $E_c$. 
Note that, in general, the system's state is represented using an ensemble $\{ ... \}$ of complex field realizations,
 each with its own set of  amplitudes $\alpha_k$. Each member breaks the gauge symmetry of a typical full quantum ensemble in a manner similar to single experimental realizations, but the ensemble preserves it \cite{Kagan97,FINESS-Book-Wright}. This naturally allows e.g. for the presence of spontaneous nonlinear many-body defects, and many interesting non-mean field phenomena that are very difficult to access using other approaches. 

For in-depth discussion of the subject, we refer the reader to \cite{FINESS-Book-Davis,FINESS-Book-Cockburn,FINESS-Book-Brewczyk,FINESS-Book-Wright} and the earlier reviews \cite{Blakie08,Brewczyk07,Proukakis08}.

%%%%%%%%%%%%%%%%%%%%%%%%%%%%%%%%%%%%%%%%%%%%%%%%%%%%%%%%%%%%%%%%%%%%%%%%%%%%%%%%%%%%%%%%%%%%%%%%%%%%%%%%%%%%%%
\subsection{Interacting 1d gas}
\label{1DGAS}
 Here, we  consider the one-dimensional Bose gas with repulsive contact interactions.
 The exact Yang-Yang solution for the uniform gas at a given interaction strength and temperature  
\cite{Yang69} will provide the benchmark to which the c-field method will be compared. 

 To obtain results independent of the trapping geometry, density profile, etc., 
 we will focus on systems that are amenable to a local density approach (LDA). 
 In this, it is assumed that the ensemble averaged density $n=\langle|\Psi(\bo{x})|^2\rangle$ varies more slowly than other relevant 
 quantities, and the system can be treated using sections of the gas of a given density $n$.
 The local density $|\Psi(\bo{x})|^2$ in a single realization can still vary strongly within the section,
 which is sufficient to include defects and turbulence phenomena.

 Working thus in the LDA with a uniform section of a larger gas,
 the grand canonical ensemble is the natural choice, as the rest of the system acts
 as a particle and thermal reservoir. 
 Results that are independent of finite-size effects are the most useful.
 Therefore we impose the thermodynamic limit by using a section with uniform density $n$ and periodic boundary conditions
 of a length $L$ sufficiently large to contain the longest length scale in the system.
 Usually this is the phase correlation length, and is equivalent to requiring the first-order (``phase'') correlation 
\eq{g1z}{
 g^{(1)}(z) = \frac{1}{n}\langle\dagop{\Psi}(x)\op{\Psi}(x+z)\rangle
}
to drop to zero when $z\gtrsim L/2$.

 For uniform systems, a basis of plane wave modes $k\equiv{\bf k}$ is natural.
 The energy cutoff for the low energy subspace $\mathcal{C}$ is equivalent to a momentum cutoff $k_c$
 so that only modes $|\mathbf{k}|<k_c$ are included. Its dimensionless form is 
\eq{fc}{
       f_c = k_c \ \frac{\Lambda_T}{2\pi} = \frac{\hbar k_c}{\sqrt{2\pi mk_BT}},
}   
in terms of the particle mass $m$ and thermal de Broglie wavelength $\Lambda_T$ at temperature $T$. The corresponding energy cutoff is 
\eq{ec}{
	\ve_c = \frac{\hbar k_c^2}{2m}=\pi\,f_c^2\,k_BT.
}

The properties of a uniform 1d gas for a given density $n$ 
 with contact interaction strength $g$
are encapsulated by two dimensionless parameters
\eq{para}{
\gamma=\frac{m g}{\hbar^2 n}; \qquad
\tau_d=\frac{T}{T_{d}} = \frac{1}{2 \pi} \frac{m k_B }{\hbar^2 } \frac{ T}{n^2}
}
(provided there are no finite-size effects).
The first parameter quantifies the interaction strength, moving from dilute Bose gases for  $\gamma\ll1$ to a strongly interacting fermionized regime when $\gamma\gg1$. 
The second one is a dimensionless temperature, in units of quantum degeneracy temperature $T_d$. 

 The parameter space of the system has been classified 
 by  the behavior of density fluctuations \cite{Kheruntsyan03,Deuar09} 
 into a number of regions separated by crossovers:
 \begin{itemize}
\item[ \textcircled{\smaller C}] Classical gas:
 $\tau_d\gtrsim 1/(4\pi)$ (physics described by classical particles, not waves).
\item[ \textcircled{\smaller D}] Quantum degenerate gas: $\sqrt{\gamma}\lesssim 4\pi \tau_d\lesssim 1$
 (Low energy modes occupied by more than one particle, density fluctuations large).
\item[ \textcircled{\smaller T}] Thermally fluctuating quasicondensate: $\gamma\lesssim 4\pi \tau_d\lesssim \sqrt{\gamma}$
 (phase coherence on appreciable scales, small density fluctuations dominated by thermal excitations, weak bunching of atoms).
\item[ \textcircled{\smaller S}] A soliton region:
  between regions \textcircled{\smaller D} and \textcircled{\smaller T} \cite{Karpiuk12,Nowicki17}.  
\item[ \textcircled{\smaller Q}] Quantum  fluctuation dominated quasicondensate: $4\pi \tau_d \lesssim \gamma \lesssim 1$
 (phase coherence on long scales, small density fluctuations dominated by quantum depletion, weak antibunching).
\item[ \textcircled{\smaller F}] Fermionized gas: $\gamma\gtrsim1$ (physics dominated by strong interparticle repulsion, overlap between single-particle wavefunctions becoming small).
\end{itemize}

A series of c-field ensembles was generated for each of many chosen parameter pairs  ($\gamma$, $\tau_d$) in parameter space. 
Each series contained separate uniform ensembles for a range of values of $f_c$. 
 We primarily used a Metropolis algorithm for a grand canonical ensemble as described in \cite{Pietraszewicz15,Witkowska10}. 
 For a few parameter pairs, in the large $\gamma$, low $\tau_d$ region where the Metropolis method was very inefficient, the ensemble was generated by 
 evolving a projected stochastic Gross-Pitaevskii equation (SPGPE) to its stationary ensemble. 
Details of our implementations are given in Appendix~\ref{S:MAINOVER}.

%%%%%%%%%%%%%%%%%%%%%%%%%%%%%%%%%%%%%%%%%%%%%%%%%%%%%%%%%%%%%%%%
\section{Observables}
\label{OBS}

 For a correct treatment of the physics of the system all the low order observables
 that form the staple of experimental measurements should be calculated correctly.
 Further, for correct dynamics, the energy and energy balance (kinetic/interaction) must also be correct.
 Any discrepancies will rapidly feed through into the other quantities in a   nonlinear system, whether in the form of energy mixing, or 
 dephasing in integrable systems. 
This argument becomes even more important 
in nonuniform systems, 
 because discrepancies in the density dependence of energy will immediately lead to bogus expansion or contraction.

 Let us define a relative discrepancy between the classical field value of an observable $\Omega^{\rm(cf)}$ obtained using a given cutoff, and the true quantum observable $\Omega$: 
\eq{rms}{
     \delta_{\Omega}(\gamma,\tau_d,f_c) := \frac{\Delta \Omega}{\Omega} =  
     \Bigg(  \frac{ \Omega^{\rm (cf) }(\gamma,\tau_d,f_c) }{ \Omega(\gamma,\tau_d)} - 1 \Bigg).
}

 We will always compare a classical field ensemble with density $n$ to exact results with the same density, so  $\delta_{n}=0$ by construction.
 This ensures  equivalence between the physical parameters $\gamma$ and $\tau_d$ \eqn{para} for both CF and exact results. 
 Next, setting $g$ and temperature $T$, leaves one free technical parameter $f_c$ for the c-field. 
 Its value can often be chosen to make one other observable match exactly, but not all.

 In an earlier study \cite{Pietraszewicz15}, the dependence of the discrepancies on cutoff was analyzed,
 and one of the most important results was that {\bf observables fall into two broad categories} described below. 

The \underline{first kind} are IR-dominated observables which display falling values with growing cutoff.
They are dominated by low energy modes and finally reach asymptotic values.
This group includes such quantities as the half-width of the $g^{(1)}(z)$ \  correlation function given by \eqn{g1z} (i.e. the phase coherence length), values of $g^{(1)}(z)$ at large distances $z$, the condensate fraction $n_0$, the effective temperature in a microcanonical ensemble \cite{Schmidt03,Davis03}, or the coarse grained density fluctuations. These last are given by 
\eq{uG}{
u_G := \frac{{\rm var} \op{N}}{\langle \op{N} \rangle} = S_0 = 
  n\, \int d{\bf z}\, {\big[} g^{\rm (2)}({\bf z})-1 {\big]} + 1
}
 and will play an important role in our analysis. \eqn{uG} is also known as the $k=0$ static structure factor $S_0$.
 It is the ratio of the measured number fluctuations to Poisson shot noise in regions that are wider than the density correlation length, and serves
 as a measure of the typical number of particles per randomly occurring density lump \cite{Pietraszewicz17b}.
 It tends: (i) to one in a coherent state or at high temperature, (ii) to zero at $T=0$ or in a fermionized gas,
 and (iii) to large positive values in the thermal-dominated quasicondensate.
 It is a density fluctuation quantity that is quite readily measured in experiments \cite{Sanner10,Muller10,Jacqmin11}
 because standard pixel resolution is sufficient, and is an intensive thermodynamic quantity.
 These features are to be contrasted to the microscopic density-density correlation
 $g^{(2)}(z) = \langle\dagop{\Psi}(x)\dagop{\Psi}(x+z)\op{\Psi}(x+z)\op{\Psi}(x)\rangle/n^2$ 
 which is usually  neither intensive nor experimentally resolvable \emph{in situ}. 

The \underline{second} category contains UV-dominated observables which follow the opposite trend.
They are underestimated for low cutoff, because high energy modes have a large contribution,
and grow with increasing $f_c$. The most prominent example are energies per particle 
\eqa{energy}{
\mc{E}_{\rm tot}&=&\mc{E}_{\rm int}+\ve\\
& =&  \frac{g}{2}\int\!dx\ \op{\Psi}^{\dagger\,2}(x)\op{\Psi}^2(x) + \int\!dx\ \frac{\hbar^2}{2m}\nabla\dagop{\Psi}(x)\nabla\op{\Psi}. \nonu
}
Others are some collective mode frequencies in a trap \cite{Bezett09a}, and $g^{(2)}(0)$.
The density $n$ also behaves this way when changing only $k_c$, but is guaranteed matched in our approach.

\begin{figure}
\begin{center}
\includegraphics[width=0.491\columnwidth]{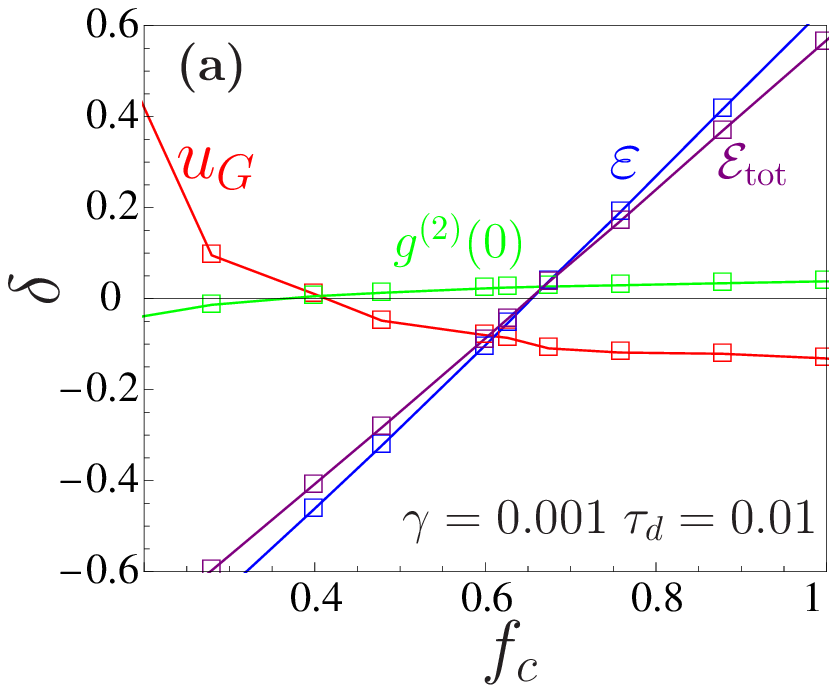}
\includegraphics[width=0.489\columnwidth]{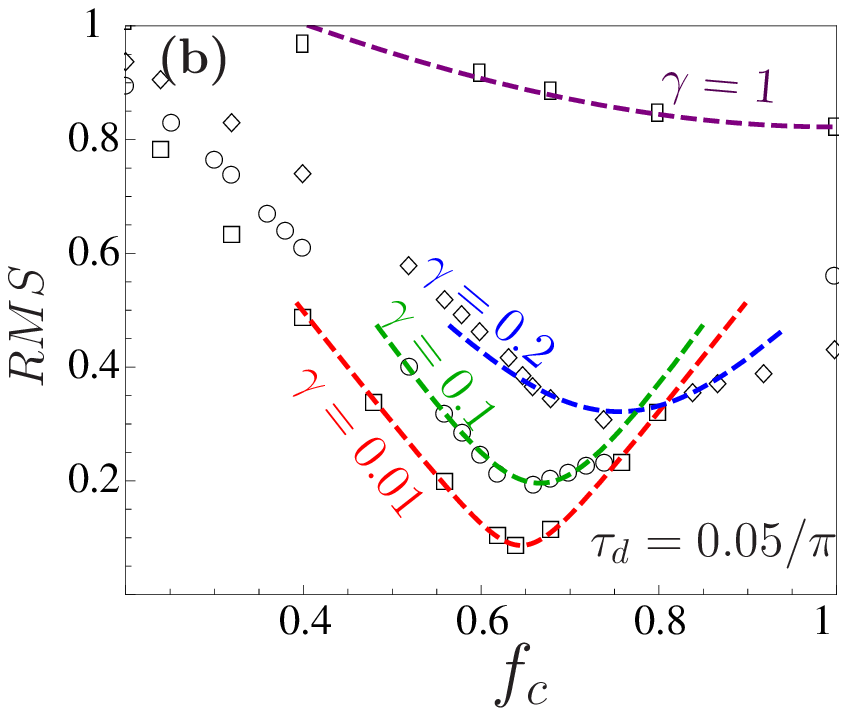}
\end{center}
\caption{
 Example of accuracy assessment for a representative choice of parameters.
 Panel (a): cutoff dependence of the discrepancies $\delta$ of single observables
 in the case $\gamma=0.001$, $\tau_d=0.01$, for
 the coarse grained density fluctuations $u_G$ (red), the total energy $\mc{E}_{\rm tot}$ (purple), the kinetic energy $\ve$ (blue),
 the density-density fluctuations, and $g^{(2)}(0)$ or the interaction energy $\mc{E}_{\rm int}$ (green). 
 Panel (b): cutoff dependence of the global discrepancy $RMS(f_c)$ at $\tau_d=0.0159$
 obtained numerically for several values of $\gamma$: 0.01 (square), 0.1 (circle), 0.2 (diamond), 1 (rectangle).
 Dashed curves show the fit (parabolic to $(RMS)^2$) to numerical points near the minimum. 
 }
\label{fig:rms-fc}
\end{figure}

 We have inspected the discrepancies \eqn{rms} for many locations in ($\gamma$,$\tau_d$) space using the numerically generated ensembles. 
 Fig.~\ref{fig:rms-fc}(a) shows a representative case for the quantum degenerate and thermally fluctuating condensate regions. 
 A variant at larger values of $\gamma$ when total energy is dominated by interaction can be found in Appendix \ref{S:fig4}.
 The exact quantum observables $\mc{E}_{\rm tot}$, $\ve$, $N$, were obtained 
 according to \cite{Yang69}, while the $g^{(2)}(0)$ according to \cite{Kheruntsyan05}. 
 The $u_G$ calculation uses a specially developed method \cite{PDJPYang17}.

 Overall, we find that the same trends seen in the 1d ideal gas are repeated in the interacting case
 for the relevant observables such as kinetic energy $\ve$ and $u_G$.  
 For the interacting gas one should also separately consider
 the interaction energy per particle $\mc{E}_{\rm int}=\tfrac{1}{2}g\,n\,g^{(2)}(0)$ and the total energy per particle $\mc{E}_{\rm tot}$.
 For the physical regimes studied, they are both found to belong to the second category: discrepancy rising with cutoff.
 In general, we can confirm that  the basic two observable categories hold also in the interacting gas.

%%%%%%%%%%%%%%%%%%%%%%%%%%%%%%%%%%%%%%%%%%%%%%%%%%%%%%%%%%%%%%%%
\section{Global accuracy indicator}
\label{PROCEDURE}

In the ideal gas, the low order observables with the most disparate behavior are the coarse-grained density fluctuations $u_G$ and kinetic energy $\ve$ \cite{Pietraszewicz15}. Their mismatch
is the strongest restriction on the range of $f_c$ for which all $\delta_{\Omega}$ errors are small.
 Based on this, a global figure of merit was defined:
$RMS_{\rm id} = \sqrt{\delta_{\ve}^2 + \delta_{u_G}^2}$.
This includes discrepancies of observables belonging to both classes.
 It had the convenient property that it was an upper bound on the error of all the observables that were treated.

 The interacting gas brings with it several additional observables. Among them, the components of energy are particularly important:
 a correct $\mc{E}_{\rm tot}$ is needed for an accurate description of  dynamics,
 $\mc{E}_{\rm int}$ for local density fluctuations, 
 while the kinetic energy $\ve$ is closely related to phase coherence length and the momentum distribution. 

 A good figure of merit $RMS(f_c)$ for the interacting case should be a bound both for $\delta_{\ve}$ and $\delta_{u_G}$ 
 as well as for $\delta_{\mc{E}_{\rm tot}}$ and $\delta_{\mc{E}_{\rm int}}$.
 Inspecting the data, one finds that the discrepancies in the total energy and interaction energy (same as for $g^{(2)}(0)$) are found to grow 
 slower than $\ve$, so that $\ve$ and $u_G$ always remain the most extreme representatives of their groups.
 However, at large interaction $\gamma$,  it is observed that near the optimum the errors
 in $\ve$ and $u_G$ can be exceeded by the error in $\mc{E}_{\rm tot}$ (see Appendix.~\ref{S:fig4}). 
 This happens when the dominant energy contribution comes from interactions, and the discrepancies of $\mc{E}_{\rm int}$ and $\mc{E}_{\rm tot}$ are about equal. 
 Hence, the overall conclusion is that $\delta_{\mc{E}_{\rm int}}$
 can be omitted from the $RMS$ without loss of generality, but  $\delta_{\mc{E}_{\rm tot}}$ should stay.

 As a result, we define the measure of global error (at a given cutoff) as:
 \eq{RMSdef}{
	RMS(\gamma,\tau_d,f_c)= \sqrt{ \Big(\,\delta_{u_G}\,\Big)^2 +
       {\rm max}\Big[ \delta^{\,2}_{\ve}, \delta_{\mc{E}_{\rm tot}}^{\,2} \Big]  }.
 }
 We use the maximum of the energy discrepancies rather than an rms of all three potentially extreme observables.
 In this form it has a more convenient interpretation, because
 (1)  we do not wish to double count the importance of energy to keep the interpretation of $RMS$
 as being close to the upper bound on the discrepancies (not $\sqrt{2}$ times the upper  bound),
 and (2) it will remain consistent with the ideal gas results of \cite{Pietraszewicz15}.
 Fig.~\ref{fig:rms-fc}(b) shows the dependence of \eqn{RMSdef} for a variety of cases. 

 The minimum of the global error quantity $RMS(f_c)$ will provide the main results in this paper.
 Its value min$RMS$, will be our figure of merit for the classical field description, 
 and its location opt$f_c$ is the globally optimal cutoff.
Namely:
 \eq{RMSdef2}{
{\rm min}_{f_c>0}\left[RMS\right] = {\rm min}RMS = RMS(\gamma,\tau_d,{\rm opt}f_c).
 }

 After generating ensembles with given values of $\gamma$, $\tau_d$, and  $f_c$, 
 observable expectation values $\Omega^{\rm(cf)}$ were calculated.
 They were compared to corresponding exact quantum values, to get the discrepancies $\delta_{\Omega}(\gamma,\tau_d,f_c)$. 
 At the end, the  $RMS(\gamma,\tau_d,f_c)$ function was minimized numerically to get min$RMS$ and opt$f_c$.
 Details of these procedures are provided in Appendix~\ref{S:rms}.

%%%%%%%%%%%%%%%%%%%%%%%%%%%%%%%%%%%%%%%%%%%%%%%%%%%%%%%%%%%%%%%%%%%%%%%%%%%%%%%%%%%%%%%%%%%%%%%%%%%%%%%%%%%%%%%%%%%%%%%%%%%%%%%%%%%%%%%%%%%%%%%%%%%%%%%%%%%%%%%%%%%%%%%%%%%%%%%%%%%%%%%%%%%%%%%%%%%%%%%%%%%%%%%%%%%%%%%%%%%%%%%%%%%%%%%%%%%%
\section{RESULTS: The Classical wave regime}
\label{RMS}

\begin{figure}[htb]
\begin{center}
\includegraphics[width=0.9\columnwidth]{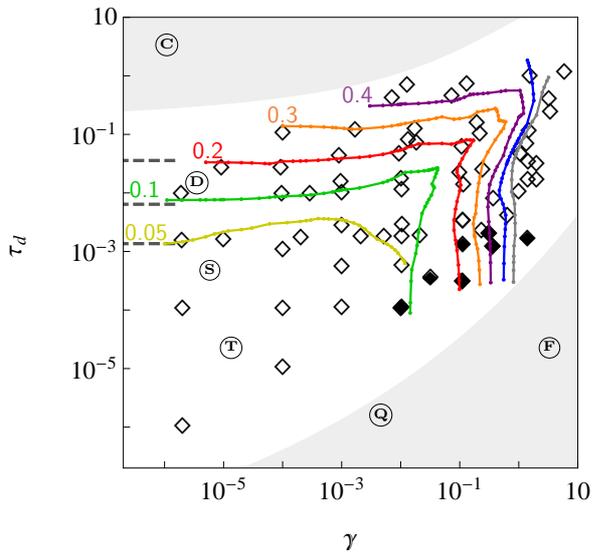}
\end{center}\vspace*{-4mm}
\caption{
The regime of applicability for classical wave fields.
Contours of min$RMS$ (the upper bound on discrepancy
of observables) are shown at values of 0.05, 0.1, 0.2, 0.3, 0.4, 0.5, and 0.6. 
The dashed lines are ideal gas positions \cite{Pietraszewicz15}.
The diagram also shows the location of numerical ensembles generated by 
Metropolis and SPGPE ensembles as open and filled diamonds, respectively.
The circled letters indicate the physical regime, as explained in the text.
}
\label{fig:minRMSpoints}
\end{figure}

 The dependence of the figure of merit min$RMS$ on physical parameters $\gamma$, $\tau_d$ is shown in Fig.~\ref{fig:minRMSpoints}. 
 This presents the \textbf{regime of applicability} of classical fields,
 depending on how much inaccuracy is to be tolerated.
 The quantity min$RMS$ bounds the inaccuracy for all observables studied.
 Open and closed points indicate for which parameters ensembles were calculated. 
 Color contours specify given values of the estimated min$RMS$.

 The white region in Fig.\ref{fig:minRMSpoints} was studied, and includes: the thermal quasicondensate \textcircled{\smaller T},
 the quantum turbulent soliton region \textcircled{\smaller S}, the degenerate gas regimes \textcircled{\smaller D}.
 The gray areas include most of the classical region \textcircled{\smaller C}, fermionized regime \textcircled{\smaller F} and the quasicondensate dominated by quantum fluctuations \textcircled{\smaller Q}. 
These are listed and delineated in Sec.~\ref{1DGAS}. 
 We have no reliable information for the gray areas because of technical difficulties in obtaining ensembles in the thermodynamic limit.
 For the coldest temperatures, especially when $\tau_d\lesssim\gamma$,  one sees
 slow convergence  during calculations  and/or increasing problems in removing finite-size effects.

 Knowing that experimental uncertainties are typically of the order of 10\%, a value of min$RMS\le0.1$ 
 tells us that the physics in this region is in practice the physics of classical matter wave fields.

 Overall, the region dominated by classical wave physics  in Fig.\ref{fig:minRMSpoints}
 is larger than one could have conservatively supposed.
 Almost the entire thermal quasicondensate \textcircled{\smaller T} is described well,  
 all the way until the crossover to fermionization kicks in around $\gamma=0.018$ to $0.075$. 
 The quoted values correspond to discrepancies min$RMS=~0.1$ and $0.2$, respectively.

 The quantum degenerate gas \textcircled{\smaller D} is correctly described for all temperatures up to at least $\tau_d \approx 0.008$ (min$RMS=0.1$) and $0.03$ (min$RMS=0.2$).
 This is an important result, as it is not immediately obvious that classical fields apply so far.
 What this means is that not only do they cover the entire \textcircled{\smaller T} regime 
 but a number of strongly fluctuating higher temperature regions as well. 
 One of them is the region with prominent thermal solitons \textcircled{\smaller S} in the range $\tau_d \sim 0.04\sqrt{\gamma}-0.2\sqrt{\gamma}$ \cite{Nowicki17,Karpiuk12}.
 At warmer temperatures\footnote{ Obtained from exact Yang-Yang calculations \cite{Yang69}.} $\tau_d\approx0.27\sqrt{\gamma}$,
 the crossover to an ideal-gas-like state that  occurs when $\mu$ changes sign
 also lies well within the classical wave region. 
 This means that the changeover from wave-like to particle-like physics occurs at much higher temperatures
 than the crossover between Bogoliubov and Hartree-Fock physics \cite{Henkel17}.

 Finally, observing the left edge of Fig.~\ref{fig:minRMSpoints},
 we can confirm that our numerical results for min$RMS$ match well to the ideal gas results of \cite{Pietraszewicz15}. 
 
 Summarizing the criterion of 10\% goodness of the c-fields description,
 the limit of the matter wave region reaches:
 (i)  $\tau_d \approx 0.008$ until $\gamma~\lesssim~0.001$,  
 (ii) $\gamma \approx 0.018$ below $\tau_d~\lesssim~0.002$.
 Between (i) and (ii), there is a bulge that extends to $\tau_d\approx\gamma\approx0.03$. 

\begin{figure}[b]
\begin{center}
\includegraphics[width=\columnwidth,clip]{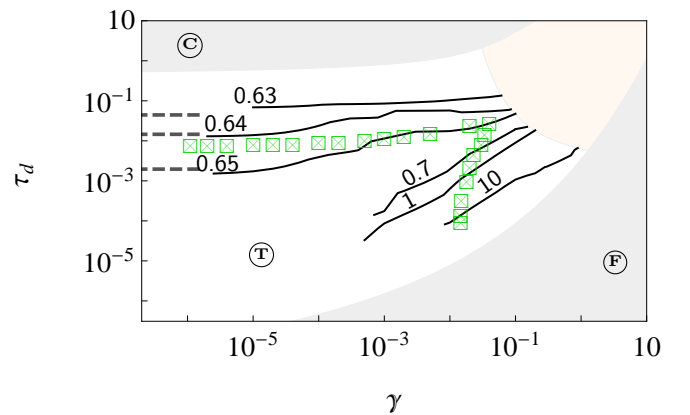}
\end{center}\vspace*{-4mm}
\caption{
Globally optimal values of the cutoff, opt$f_c$ (black contours with values shown on the plot). 
For reference, green symbols show the location of min$RMS=0.1$ from Fig.~\ref{fig:minRMSpoints}. 
The salmon colored area indicates a region 
in which there was  insufficient precision in the numerical ensembles to determine the position of the closely spaced contours.
The dashed gray lines are the ideal gas predictions for opt$f_c$. 
\label{fig:optfc}}
\end{figure}

 Together with  min$RMS$, also the \textbf{optimal cutoff} opt$f_c$ was obtained. 
 Its behavior is presented in Fig.\ref{fig:optfc}.
 Black contours specify given values of opt$f_c$
 estimated from the Metropolis/SPGPE data as described in Appendix.~\ref{S:CONTOUR}. The green points are a copy of the min$RMS=0.1$ location.
 For higher temperatures the opt$f_c$ values are changing slowly between contours.
 This behavior specifies a flat region in which 
 a constant opt$f_c\approx 0.64$ is a very good approximation.

 Around $\tau\approx 0.1 \gamma$ a change in behavior occurs 
 and the cutoff begins to grow at very low-temperatures, towards the \textcircled{\smaller F} regime.
 This is evidenced by the large jump from values of $1$ to $10$ 
 on a similar contour spacing as between  $0.65$ and $0.64$.  
 This opt$f_c$ growth  corresponds also to a crossover from the thermal to quantum fluctuation dominated quasicondensate, 
 and to behavior like in Fig.~\ref{sfig:rms} in Appendix~\ref{S:fig4}.

All in all, there appear two regions (on either side of $\tau_d\sim 0.1 \gamma$)
 in which the dependence of opt$f_c$ on $\gamma$ and $\tau_d$
is either flat or significant.
 In the flat region dominated by thermal fluctuations 
 (up to min$RMS\lesssim0.2$) one can state that:
\begin{subequations}\label{constfc}
\eq{constfc-a}{
{\rm opt}f_c=0.64\pm0.01
}
i.e.
\eq{constfc-b}{
k_c=(1.60\pm0.03)\frac{\sqrt{mk_BT}}{\hbar};\quad \ve_c = (1.29\pm0.04)\,k_BT .
}
\end{subequations}

Confidence in our results is added by the fact that
they are consistent with the ideal gas from \cite{Pietraszewicz15}
which reported  opt$f_c\to\zeta(3/2)/4=0.653$ till $\tau_d\le 0.00159$. 
The case of low $\gamma$ 
($\gamma<0.005$ in Fig.~\ref{fig:optfc}) agrees with this.

%%%%%%%%%%%%%%%%%%%%%%%%%%%%%%%%%%%%%%%%%%%%%%%%%%%%%%%%%%%%%%%%%%%%%%%%%%%%%%%%%%%%%%%%%%%%%%%%%%%%%%%%%%%%%%
\subsection{Analytical approximations for opt$f_c$ and min$RMS$}

\label{APPROX}

 Some useful analytical approximations regarding opt$f_c$ and min$RMS$ are readily obtained
 for the flat opt$f_c$ region using the ideal gas. 
 From  \cite{Pietraszewicz15} we know that 
the discrepancy in $u_G$ becomes very flat  for a wide range of $f_c$, 
while discrepancy in $\ve$ remains always highly sensitive. 
In fact, this kind of behavior is typical for many parameters, not just the ideal gas (see Fig.~\ref{fig:rms-fc}(a)). 

The quantum prediction for $\ve$ in the  thermodynamic limit $L\to\infty$ at $\gamma=0$ \quad is \quad
$\ve^{({\rm q})} = \frac{1}{2\pi n}\int_{-\infty}^{\infty} \!dk\,\frac{\hbar^2k^2}{2m}\,n^{({\rm q})}_k$
with Bose-Einstein mode occupations 
\eq{occup}{
n^{({\rm q})}_k = [e^{(\hbar^2k^2/2m-\mu)/k_BT}-1]^{-1},
} 
while the corresponding
 c-field quantity $\ve^{\rm(cf)}$ uses the cutoff $\int_{-k_c}^{k_c}$ and Rayleigh-Jeans occupations from equipartition: 
\eq{occupcf}{
n^{({\rm cf})}_k = k_BT/\big[\tfrac{\hbar^2k^2}{2m}-\mu\big].
}
Now, simple calculations lead to:
\eq{deltaid}{
\delta_{\ve} = \frac{4f_c}{\zeta(\tfrac{3}{2})}-1 + \mc{O}(\tau_d^{\frac{1}{2}});\quad
\delta_{u_G} = \sqrt{\tau_d}\,\left[\frac{6}{f_c\pi}+3\zeta(\tfrac{1}{2})\right] + \mc{O}(\tau_d)
}
Analysis of \eqn{deltaid} shows that no positive $f_c$ value can satisfy $\delta_{u_G}=0$. 
In effect, the optimal cutoff value, i.e. 
\eq{fc065}{
{\rm opt}f_c \approx \frac{\zeta(\tfrac{3}{2})}{4} \approx 0.653
}
is set only by the location of $\delta_{\ve}=0$. 
In turn, from the $\delta_{u_G}$ expression one has an estimate of 
\eq{minrmsest}{
{\rm min}RMS = \sqrt{\tau_d}\left|\frac{24}{\pi\zeta(\tfrac{3}{2})}+3\zeta(\tfrac{1}{2})\right| \approx 1.46\,\sqrt{\tau_d}.
}
The estimated values of $\tau_d=0.0047$ and $0.019$ for the location of min$RMS=0.1$ and 0.2, respectively, are relatively close to the actual numerical values.

%%%%%%%%%%%%%%%%%%%%%%%%%%%%%%%%%%%%%%%%%%%%%%%%%%%%%%%%%%%%%%%%%%%%%%%%%%%%%%%%%%%%%%%%%%%%%%%%%%%%%%%%%%%%%%%%%%%%%%%%%%%%%%%%%%%%%%%%%%%%%%%%%%%%%%%%%%%%%%%%%%%%%%%%%%%%%%%%%%%%%%%%%%%%%%%%%%%%%%%%%%%%%%%%%%%%%%%%%%%%%%%%%%%%%%%%%%%%

\section{ Beyond the uniform gas}
\label{ADDINFO}
%%%%%%%%%%%%%%%%%%%%%%%%%%%%%%%%%%%%%%%%%%%%%%%%%%%%%%%%%%%%%%%%%%%%%%%%%%%%%%%%%%%%%%%%%%%%%%%%%%%%%%%%%%%%%%
\subsection{Nonuniform density }
\label{NONUNIFORM}

 In a nonuniform gas, neighboring sections described by the LDA have different densities,
 so results assigned for them are placed differently
 on Figs.~\ref{fig:minRMSpoints}-\ref{fig:optfc}.
 In thermal equilibrium they follow the line $\tau_d(n) \propto \gamma^2 $.
 The central core of the gas cloud is not usually in the quantum fluctuation region \textcircled{\smaller Q} because 
 cooling that far is difficult.
 So then, the central core is in the constant opt$f_c\approx0.64$ region,
 and so are the remaining gas sections, apart for the dilute tails.
 This means that all the sections can be described together using the same common cutoff $k_c$ on a common grid. 

 A further conjecture under these conditions would be that a good description 
 can be obtained regardless of whether a plane wave or a harmonic oscillator basis is chosen
 (at least if one uses opt$f_c$). The agreement seen between numerical studies of trapped gases using plane wave bases and experiment 
\cite{Cockburn11c,Karpiuk12,Cockburn12} corroborates  the above statement.

\subsection{Comparison to past results}
\label{NC}

 In the history of  the subject, the cutoff has been given either
 in terms of the highest single-particle energy $E_c$ (equivalent to the wavevector $k_c$),
 or in terms of the c-field occupation $N_c=\langle|\alpha_{|k|=k_c}|^2\rangle$ of the cutoff mode.

 The relationship between $N_c$ and $E_c$ is the following:
 \eq{Ncid}{
 N_c \approx \frac{k_BT}{E_c}. 
 }
 Our finding of optimal cutoff opt$f_c=0.64$  
 indicates $N_c =~ \tfrac{1}{\pi ({\rm opt}f_c)^2}=~0.78$ in 1d.

 How does this compare to other studies? 
 The fundamental observation in the early work  was that 
 it should occur somewhere around $N_c\sim\mc{O}(1-10)$ \cite{Davis01a,Schmidt03}.
 Since the focus was mainly on qualitative results, cutoffs were simply postulated.
 This has also been done in later approaches which used $N_c=1$ \cite{Brewczyk04} or $N_c=2$ \cite{Rooney10}.

 The authors of \cite{Zawitkowski04,Brewczyk07}  
  obtained $N_c=~0.6~-~0.7$  in 3d by matching condensate fraction $n_0$ to the ideal gas value.
 Although a 40\% discrepancy in $\ve$ arose, getting $n_0$ correct was considered much more important.
 In turn, others found that changing the cutoff value of $E_c$ by 20\%
 introduced only a few percent difference in $n_0$ \cite{Bradley08}. 
 
 In turn, the authors of \cite{Witkowska09} optimized the cutoff to minimize the mismatch in the full distribution of the excited fraction in an ideal gas. 
 They find $N_c=1$ (hence opt$f_c=0.56$) in a 1d trap in the canonical ensemble,  
  whose local LDA density segments are comparable to the treatment here. 
 The cutoff obtained by \cite{Witkowska09} was used for weakly interacting gases by \cite{Bienias11a,Bienias11b}
 who found a good match for  condensate fraction fluctuations,   
 but 10\% discrepancies in $g^{(2)}(0)$.

 Most previous cutoff determinations were based on optimizing single, IR-dominated observables.
 In contrast, we also include observables from the UV-dominated group and 
 show that the discrepancy $\delta_{\ve}$ tends to be more sensitive to $f_c$ than the other $\delta_{\Omega}$.
 What was not noted before is that the modes that contribute most 
 to $\ve$ are not the same as for the majority of the other observables. 
 Hence, raising the cutoff in energy to opt$f_c$ does not significantly affect the observables of the first kind.

%%%%%%%%%%%%%%%%%%%%%%%%%%%%%%%%%%%%%%%%%%%%%%%%%%%%%%%%%%%%%%%%%%%%%
\subsection{Relevance of high cutoff for dynamics}
\label{DYNAMICS}

 The lower cutoffs (like those proposed in the past) generate a large error in kinetic energy.
  This can be problematic for nonlinear dynamics, where correct energies become crucial.
 Any errors in energy in one part of the system rapidly infect the rest  
 with errors through the nonlinearity, and for example lead to spurious movement of mass. 

 An example of dynamics that is adversely affected are collective mode frequencies,
 studied by experiment \cite{Jin97}, c-fields, and other theory. 
 The experiment determined a rapid increase in the frequency of the $m=0$ quadrupole mode
 from about $1.85\omega_{\perp}$ below $T\approx 0.6 T_c$ to $2\omega_{\perp}$ above $T\approx 0.7 T_c$.  
 The ZNG theory \cite{Jackson02} and the second-order Bogoliubov of Morgan\etal \cite{Morgan03}
 fairly well matched this phenomenon. 
 In contrast, c-field calculations  \cite{Bezett09a} with lower opt$f_c$ did not predict a rise in frequency.  
 The disagreement was attributed to inadequate description of the dynamics of the above-cutoff modes.
 
 Interestingly, in \cite{Bezett09a} 
 the relevant frequency rose up to $1.9\omega_{\perp}$ at the highest 3d cutoff 
 when 65\% of the atoms were in the c-field. The cutoff increase was not taken further
 due to anxiety over including poorly described modes with small occupation. 
 Later, \cite{Karpiuk10} used an even higher energy cutoff and recovered the frequency increase 
 but at an excessive temperature $T\approx0.8T_c$. 
 
 Our present result showing a weak cutoff dependence of non-kinetic observables suggests that 
 with an even higher cutoff, such as opt$f_c$, further improvement
 in the collective mode's description may be possible. 
 There would also be little detrimental effect  on condensate fraction.

%%%%%%%%%%%%%%%%%%%%%%%%%%%%%%%%%%%%%%%%%%%%%%%%%%%%%%%%%%%%%%%%%%%%%%%%%%%%%%%%%%%%%%%%%%%%%%%%%%%%%%%%%%%%%%%%%%%%%%%%%%%%%%%%%%%%%%%%%%%%%%%%%%%%%%%%%%%%%%%%%%%%%%%%%%%%%%%%%%%%%%%%%%%%%%%%%%%%%%%%%%%%%%%%%%%%%%%%%%%%%%%%%%%%%%%%%%%%

\section{Wrap-up}

\label{CONCLUSIONS}

 We have determined the region of parameter space
 of the 1d interacting Bose gas in which matter
 wave physics applies. It is shown in Fig.~\ref{fig:minRMSpoints}.  
 For 10\% error in standard observables or less, the limits lie at $\gamma=0.018$ and $\tau_d=0.008$ (i.e. $\tau=0.1$ in the notation of \cite{Kheruntsyan03}),
 with an additional bulge extending somewhat beyond these.
 We claim that quantitatively accurate studies can be confidently carried out with the classical field approach provided the system stays in this region, i.e. in the thermal quasicondensate, the quantum turbulent regime, and the whole crossover into Hartree-Fock physics.  
 
 The appropriate cutoff choice to capture the behavior of many observables 
 simultaneously and obtain the correct energy in the system 
 is shown in Fig.~\ref{fig:optfc}. 
 For $k_BT\gtrsim\mu$, when the fluctuations are thermally dominated, opt$f_c$ is uniformly $\approx0.64$.
 This justifies the use of a single cutoff for plane-wave modes even for inhomogeneous gases
 and suggests that different bases are equivalent in this regime.
 
 The globally optimal cutoff that we obtain is noticeably higher in energy than most prior determinations. 
 However this result is still consistent with earlier determinations because
 when min$RMS$ remains small, the discrepancies in the various observables remain small as well.
 The goodness of the c-field description depends on how large min$RMS$ is.

 An important observation is that the energy per particle (especially kinetic)
 is the most sensitive criterion for a correct overall description in the interacting gas.
 This is especially relevant for strongly nonstationary dynamics where errors in energy feed through into errors in dynamics.
 As a result, the best looking option for a consistent system description are c-fields with a high cutoff.

 Finally, the approach used here could also work in attractive, or multicomponent gases
 and in higher dimensional systems, where the cutoff dependence
 could be markedly different \cite{Witkowska09,Pietraszewicz15}.
 It should also shed light on the question of whether 
 the inconsistency of c-fields in the 2d ideal gas identified in \cite{Pietraszewicz15}
 abates when interactions are present.
 Simultaneously accurate values of $u_G$ and kinetic energy were impossible to obtain even in the limit
 $T\to0$. 
 The greatest difficulty for such a study is how to obtain accurate 2d/3d results
 to compare with the c-field description. In the high-temperature region, Hartree-Fock methods
 can be exploited for this purpose, because as temperature rises they eventually become more accurate
 than classical fields \cite{Henkel17}.

 A topic for a future paper will concern the fermionized and
 quantum fluctuating quasicondensate at lower temperatures. We were not able to reach these regimes here,
 but they can be accessed with the extended Bogoliubov theory
 of Mora \& Castin \cite{Mora03}.

%%%%%%%%%%%%%%%%%%%%%%%%%%%%%%%%%%%%%%%%%%%%%%%%%%%%%%%%%%%%%%%%%%%%%
%Acknowledgments
%%%%%%%%%%%%%%%%%%%%%%%%%%%%%%%%%%%%%%%%%%%%%%%%%%%%%%%%%%%%%%%%%%%%%
\acknowledgments
We are grateful to Mariusz Gajda, Matthew Davis, Blair Blakie, and Nick Proukakis
for helpful discussions, and Isabelle Bouchoule for insight into her experiment which showed the importance of $u_G$ \cite{Jacqmin11}. 
This work was supported by the National Science Centre grant No. 2012/07/E/ST2/01389.

\bibliography{cfields}

\appendix

\section{Numerical classical field ensembles}
\label{S:MAINOVER}

%%%%%%%%%%%%%%%%%%%%
\subsection{Overview }
\label{S:OVER}
For each of the pairs of parameters $\gamma,\tau_d$ shown in Fig.~\ref{fig:minRMSpoints}, about 10-15 ensembles of $\Psi(x)$ with different cutoff values $f_c$ were generated and used to evaluate c-field observables $\Omega^{\rm(cf)}(\gamma,\tau_d,f_c)$. 
The values of $f_c$ were chosen individually for each $\gamma,\tau_d$ pair to cover the region of the minimum of $RMS(f_c)$ with a resolution that is sufficient to determine min$RMS$ and opt$f_c$ to a satisfactory degree. This was usually 10-20 values, as shown e.g. by the data points in Fig.~\ref{fig:rms-fc}. 
Each ensemble consisted of $10^3-10^4$ members. The members were generated using either a Metropolis algorithm (see Sec.~\ref{S:METRO}) or a projected stochastic Gross-Pitaevskii (SPGPE) equation simulation (see Sec.~\ref{S:SGPE}). The former are shown as empty symbols in Fig.~\ref{fig:minRMSpoints} and the latter as filled symbols.

 The generation of the ensembles is parametrized by the values of $T$, $\mu$, $g$, and the numerical lattice.
 This is not immediately convertible to $\gamma$ and $\tau_d$, since these  quantities depend on $n$,
 and $n=\langle N\rangle/L$ depends numerically on the actual ensemble generated.
 To deal with this inverse problem, we proceeded as follows: First a target pair of $\gamma^{\rm target},\tau_d^{\rm target}$ is chosen.
 This gives $g$ and a target density $n^{\rm target}$ from \eqn{para}.
 Next, the Yang-Yang exact solution \cite{Yang69}
 is obtained, giving the value of $\mu$ that generates the target density in the full quantum description\footnote{
 There is one free scaling parameter in the description using $\gamma$ and $\tau_d$, which we set in the Yang-Yang calculations using the arbitrary choice $k_BT=1$.}.
 This is usually close but not exactly equal to a chemical potential $\mu^{\rm(cf)}(f_c)$ that would give the same density of the c-field ensemble.
 In any case, $\mu^{\rm(cf)}$ depends on the chosen $f_c$. Nevertheless, since $\mu$ and $\mu^{\rm(cf)}$ are close,
 we simply use the target quantum $\mu$ and various values of $f_c$ to generate each ensemble, knowing that it will be close to $\gamma^{\rm target}$ and $\tau_d^{\rm target}$.
 
 Each ensemble with a different $f_c$ generates a slightly different mean particle density $n^{\rm(cf)}(f_c)=\langle N\rangle/L$,
 and corresponding $\gamma^{\rm(cf)}(f_c)$,   $\tau_d^{\rm(cf)}(f_c)$
 which lie close to but not exactly at $\gamma^{\rm target},\tau_d^{\rm target}$.
 However, it is not necessary to hit exact predetermined target values for our purpose of generating contour diagrams. 
 Instead, the value of $n$ at the optimal cutoff opt$f_c$ was the one used to determine the operational values of 
\eq{gamtau}{
\gamma=\frac{mg}{\hbar^2n^{\rm(cf)}(\small{\rm opt}f_c)},\qquad
\tau_d=\frac{mk_BT}{2\pi\hbar^2 n^{\rm(cf)}({\rm opt}f_c)^2}
}
 used for the analysis (and shown in Fig.~\ref{fig:minRMSpoints}). 

 The numerical lattice itself, is chosen according to the usual criteria to obtain a system in the thermodynamic limit.
 The box length $L$ must be sufficient to capture the longest length scales.
 The longest feature is the width of the $g^{(1)}(z)$ phase correlation function,
 and $L$ was chosen so that $g^{(1)}$ decays to zero before reaching a distance of $z=L/2$. 
 Namely,  $g^{(1)}(L/2)$ found from the ensemble falls closer to zero than its statistical uncertainty.
 On the other hand, the lattice spacing $\Delta x$ must be sufficiently small to resolve the smallest allowable features.
 These are the density ripples of a standing wave composed of waves with the cutoff momentum $k_c$.
 That is we need $\Delta x\le\pi/(2k_c)$. In practice we took a several times finer spacing $\Delta x$
 to smooth the visible features. The numerical lattices contained $M=2^{10} - 2^{12}$ points.

 Periodic boundary conditions were used, to have easy access to plane wave modes through Fourier transforms. 
 The c-field was given support only within the low-energy subspace $\mc{C}$ by keeping only
 the $M_{\mc{C}}$ plane wave components of the Fourier transformed field with $|k|\le k_c$.

%%%%%%%%%%%%%%%%%%%%
\subsection{Metropolis algorithm}
\label{S:METRO}

Our application of the Metropolis method to generate c-field ensembles follows the approach of Witkowska\etal\ \cite{Witkowska10}, with minor modifications as used in \cite{Pietraszewicz15}. The latter paper introduced amendments to generate grand canonical ensembles, primarily by removing the conservation of $N$ used in \cite{Witkowska10}. This has the additional advantage of removing the need for a small but tricky compensation that is otherwise needed to preserve detailed balance in the number-conserving case. 

We aim to generate the grand canonical probability distribution
\eq{GCE}{
P(\Psi) \propto \exp\left[-\frac{E_{\rm kin}(\Psi)+E_{\rm int}(\Psi)-\mu N(\Psi)}{k_BT}\right].
}
where 
\eq{Ncf}{
N(\Psi) = \sum_x \Delta x |\Psi(x)|^2,
}
and energies are
\eqa{Ecf}{
E_{\rm int}(\Psi) &=& \frac{g\Delta x}{2}\sum_x |\Psi(x)|^4,\\
E_{\rm kin}(\Psi) &=& \frac{\hbar}{2m}\sum_k \Delta k\, k^2\,|\wt{\Psi}(k)|^2.
}
The kinetic energy uses the Fourier transformed field normalized to make $\sum_k \Delta k\, |\wt{\Psi}(k)|^2= N$, i.e. 
\eq{FT}{
\wt{\Psi}(k) = \frac{1}{\sqrt{2\pi}}\sum_x \Delta x\, e^{-ikx} \Psi(x)
}
with wavevectors $k = 2\pi j/L=j\Delta k$, and $j$ integers. 

The starting state was $\Psi_0(x) = 0.$ 

A random walk is then initiated which generates a Markov chain with members $\Psi_s(x)$ after each step $s$.
A trial update $\Psi^{\rm trial}(x)$ is generated at each step.
The ratio of probabilities 
\eq{rr}{
r = \frac{P(\Psi^{\rm trial})}{P(\Psi_s)}
} 
is evaluated. The update is accepted always if $r>1$, or with probability $r$ if $0<r<1$. Then  the next member of the random walk $\Psi_{s+1}$ becomes $\Psi^{\rm trial}$.
Otherwise the update is rejected, and $\Psi_{s+1}=\Psi_s$.

We used two kinds of updates, chosen randomly at each step:
\begin{enumerate}
\item $99\%/M_{\mc{C}}$ probability:
A change of the amplitude of one of the plane wave modes $k'$, such that $\wt{\Psi}^{\rm trial}(k') = \wt{\Psi}_s(k')+\delta$ while the other modes are unchanged: 
$\wt{\Psi}^{\rm trial}(k\neq k') = \wt{\Psi}_s(k)$. The random shift $\delta$ is a Gaussian distributed complex random number with amplitude chosen so that the acceptance ratio is about 50\%.
The value of $k'$ to change is chosen randomly from the $M_{\mc{C}}$ plane wave modes that lie below the energy cutoff. 
\item 1\% probability: We found that it is necessary to sometimes slightly shift the center of mass of the system to  break the system out of getting stuck on a nonzero mean velocity. 
For this, one generates $\wt{\Psi}^{\rm trial}(k) = \wt{\Psi}_s(k\pm\Delta k)$, shifting all values by one lattice point.
The sign is chosen randomly, and the wavefunction $\Psi(\wb{k})$ at the marginal value of $\wb{k}=\pm(k_c+\Delta k)$ that overflows $k_c$ due to the shift is moved to the opposite end of the spectrum at $\mp(k_c)$.
\end{enumerate}
Since all of these updates preserve detailed balance individually, no additional compensation to $r$ is required to determine acceptance, unlike in the $N$-conserving algorithm for the canonical ensemble. 

The correlation of various quantities such as $E$, $N$, $g^{(1)}(z)$ and the center-of-mass momentum $k_{COM}$ over subsequent steps $s$ is tracked. 
Ergodicity is then exploited to obtain independent ensemble members by placing only every $\Delta s$-th member of the Markov chain into the final ensemble. The spacing $\Delta s$ is chosen sufficiently large to make all the observables uncorrelated. Also, the first $t_s\gtrsim\Delta s$ elements of the Markov chain are discarded to allow for the dissipation of transients caused by the starting state. 
The required $\Delta s$ and $t_s$ depend on the regime studied. Generally speaking, when  $\gamma<\tau_d$, we had $\Delta s = \mc{O}(10^4-10^5)$, while in the 
opposite regime ($\gamma>\tau_d$) $\Delta s$ was even larger, causing us to switch to using the SPGPE algorithm.

%%%%%%%%%%%%%%%%%%%%
\subsection{SPGPE algorithm}
\label{S:SGPE}
The projected stochastic Gross-Pitaevskii equation (SPGPE) \cite{Gardiner03,Bradley14} is here
\eqa{SPGPE}{
i\hbar\frac{d\Psi(x)}{dt} &=& (1-i\gamma_{\mc{C}})\mc{P}_{\mc{C}}\Bigg\{\left[-\frac{\hbar^2}{2m}\frac{d^2}{dx^2} -\mu + g|\Psi(x)|^2\right]\Psi(x) \nonu\\
&&+ \sqrt{2\hbar\gamma_{\mc{C}}k_BT}\,\eta(x,t).\Bigg\}
}
This approach has been extensively used to generate grand canonical ensembles of $\Psi(x)$ and described in detail in \cite{Duine01,Gardiner03,Proukakis08,FINESS-Book-Cockburn}. The quantity $\eta(x,t)$ is a complex time-dependent white noise field with the properties
\eqs{eta}{
\langle\eta(x,t)\eta(x',t')\rangle &=& \langle\eta(x,t)\rangle = 0\\
\langle\eta(x,t)^*\eta(x',t')\rangle &=& \delta(x-x')\delta(t-t').
}
It is generated using Gaussian random numbers of variance $1/(\Delta x\Delta t)$  
where time steps are $\Delta t$ and the numerical spatial lattice is $\Delta x$.
The $\mc{P}_{\mc{C}}$ is a projector onto the low-energy subspace.
A quantity $\mc{P}_{\mc{C}}\{A(x)\}$ is implemented by Fourier transforming $A(x)$ to k-space, zeroing out all components with $|k|>k_c$, and Fourier transforming back again to x-space.

The physical model in the SPGPE treats the above-cutoff atoms as a thermal and diffusive bath with temperature $T$, chemical potential $\mu$, and a coupling strength $\gamma_{\mc{C}}$ between the low- and high-energy subspaces. We typically used $\gamma_{\mc{C}}=0.02$. The long-time limit of an ensemble of many such trajectories is the grand canonical ensemble with $T$ and $\mu$ the same as the bath. To obtain an ensemble we started with the standard vacuum initial states $\Psi(x)=0$ and ran the simulation multiple times, with new noises in each run to generate a new trajectory.  
Ensemble averaged quantities were tracked over time until all reached stationary values. Then, the fields at this stabilized time were taken as the members of the final ensemble, usually with 400-2000 members.

Apart from faster convergence, this approach requires less numerical tweaking than the Metropolis algorithm since it is unnecessary to search for the correlation time $\Delta s$ in the Markov chain,  
while the size of the fluctuations is chosen automatically instead of optimizing $\delta$ to get reasonable acceptance rates. On the other hand, only the grand canonical ensemble can be generated using \eqn{SPGPE}. 
However, a more complicated ``scattering term'' SPGPE  derived by Rooney \cite{Rooney12}
does conserve $N$, and a simple modified SPGPE for canonical ensembles has been derived recently \cite{Pietraszewicz17b}.

%%%%%%%%%%%
\subsection{Discrepancies at lower temperature}
\label{S:fig4}

\begin{figure}[htb]
\begin{center}
\includegraphics[width=0.8\columnwidth]{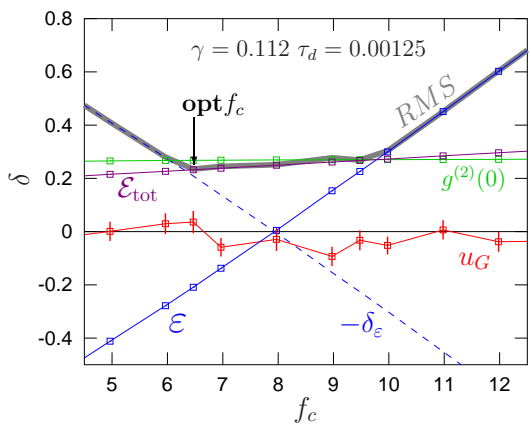}
\end{center}\vspace*{-4mm}
\caption{
Cutoff dependence of the discrepancies $\delta$ of single observables for the case $\gamma = 0.112,
\tau_d = 0.00125$. Notation as in Fig.~\ref{fig:rms-fc}(a). The symbols show results obtained numerically with the SPGPE \eqn{SPGPE}. 
The figure of merit $RMS(f_c)$ is shown as the thick grey line.
Error bars are shown for $\delta_{u_G}$, while the remaining error bars are below resolution. The dashed blue line shows $-\delta_{\ve}$ as a  reference.
The arrow indicates the choice of the optimal cutoff.  
}
\label{sfig:rms}
\end{figure}

Fig.~\ref{sfig:rms} shows cutoff dependence for a representative case of large interaction and low temperature
 that  was calculated with the SPGPE.

%%%%%%%%%%%%%%%%%%%%%%%%%%%%%%
\section{ Optimization procedures}
\label{S:NUMERIC2}

%%%%%%%%%%%
\subsection{Calculation of optimal cutoff, figure of merit } 
\label{S:rms}

For numerically generated ensembles, the set of data we use has been described in Sec.~\ref{S:OVER}:
 about 10-15 c-field ensembles of $\Psi(x)$ generated for different cutoff values $f_c$
 but the same $\mu$, as shown in Fig.~\ref{fig:rms-fc}(a).
 The target was to obtain a resolution that is sufficient to determine min$RMS$ and opt$f_c$ to a satisfactory degree. 
 
 To match these ensembles, exact quantum results were obtained using the Yang-Yang theory, but now with a
 chemical potential chosen individually for each $f_c$ to obtain the same density as $n^{\rm(cf)}(f_c)$.
 Observable discrepancies $\delta_{\Omega}(f_c)$ were calculated according to \eqn{rms}.
 Then the values of $RMS(f_c)$ were calculated using \eqn{RMSdef}.
 All these were inspected on plots like those shown in Fig.~\ref{fig:rms-fc} and \ref{sfig:rms}. 
 
 The approach taken to identify opt$f_c$ depends now on the behavior of the minimum of $RMS(f_c)$. 

 A \emph{rounded minimum} occurs when the interaction energy has only a minor contribution near opt$f_c$. Behavior of this kind is seen in Fig.~\ref{fig:rms-fc}.
 It happens typically in the hotter part of parameter space when thermal fluctuations dominate and $\gamma$ is relatively small 
(it was already seen in the ideal gas \cite{Pietraszewicz15}). 
Both $\delta_{u_G}$ and $\delta_{\ve}$ are approximately linear  in the region near the minimum of $RMS$, whereas $\delta_{\mc{E}_{\rm tot}}\approx\delta_{\ve}$. Then $(RMS)^2$ is given approximately by a sum of two parabolas --- i.e. a parabola in $f_c$. 
Accordingly, we make a least squares fit of the numerical values of $RMS$ to the parabola 
\eq{RMSfit}{
RMS(f_c)^2 = a\left(f_c-{\rm opt}f_c\right)^2+({\rm min}RMS)^2.
}
with fitting parameters opt$f_c$, min$RMS$ and $a$  in the vicinity of the minimum. Examples are seen in Fig.~\ref{fig:rms-fc}(b). The fitted opt$f_c$ and min$RMS$ are our final estimates.

 The other possibility is a \emph{flat-bottomed minimum} like that shown in Fig.~\ref{sfig:rms}.
 This occurs when the error in $g^{(2)}(0)$ (and by implication in $\mc{E}_{\rm int}$ and $\mc{E}_{\rm tot}$)
 is sufficiently large near the minimum to compete with or exceed the error in $u_G$.
 In the region in which $|\delta_{\ve}|\le|\delta_{\mc{E}_{\rm tot}}|$, the error in $\ve$ becomes negligible,
 and so $RMS=\sqrt{\delta_{u_G}^2+\delta_{\mc{E}_{\rm tot}}}$. Both of these two remaining errors usually depend weakly on $f_c$. 
 In this case, we choose the smaller of the two $RMS$ values at the ``corners'' 
 of the flat-bottomed minimum that occur at $\delta_{\ve}=\delta_{\mc{E}_{\rm tot}}$ as min$RMS$, and the corresponding value of $f_c$ as opt$f_c$.
 Some rare intermediate cases, such as when the maximum error swaps between $\delta_{u_G}$ and $\delta_{\mc{E}_{\rm tot}}$
 within the flat-bottomed minimum are dealt with on a case-by-case basis after inspection of the plot of $RMS(f_c)$.

%%%%%%%%%%%
\subsection{Generation of contour lines} 
\label{S:CONTOUR}

The contours in Figs.~\ref{fig:minRMSpoints} and~\ref{fig:optfc} were obtained using Lagrangian interpolation of a function of two variables.
Triangular polygons are successively selected using three corner points (labeled below as $i=1,2,3$), chosen from among those shown in Fig.~\ref{fig:minRMSpoints}.	 
Within such a triangle, the interpolation of a function $F$ that takes values $F_i$ at the corner points is given by 
\eq{LI1}{
F(\gamma,\tau_d)=\sum_j N_i(\gamma,\tau_d)\, F_i.
}
Here
\eq{LT2}{
N_i(\gamma,\tau_d) = \frac{1}{2A}\left(\alpha_i+\beta_i \gamma+\zeta_i \tau_d\right),
}
$A$ is the area of the triangle, and the coefficients are
\eqa{LT3}{
\alpha_i &=& \gamma_{i+1}\tau_{d,i+2}- \gamma_{i+2}\tau_{d,i+1}\nonu\\
\beta_i &=& \tau_{d,i+1}-\tau_{d,i+2}\\
\zeta_i &=& \gamma_{i+2}-\gamma_{i+1}\nonu
}
where the indices $i+1$ and $i+2$ are understood as being modulo 3.
The locus of a contour is obtained by requiring a given value of $F$ (e.g. $F=0.1$) at chosen locations.

A precise determination of the behavior of opt$f_c$ in the light orange colored region
 of Fig.~\ref{fig:optfc} proved elusive, however. 
 Much larger numerical ensembles than those we generated would be necessary to get higher precision,
 as well as a finer spacing in $\gamma$,$\tau_d$ parameter space.
 It seems that this light orange region is a little practical importance since min$RMS$ becomes large there and a c-field treatment is no longer recommended.

\end{document}